\documentclass[journal, letterpaper, twocolumn]{IEEEtran}

\usepackage[T1]{fontenc}
\usepackage[latin9]{inputenc}
\usepackage{algorithm}
\usepackage{amsmath}
\usepackage{amssymb}
\usepackage{graphicx}
\usepackage{setspace}
\usepackage{bm}
\usepackage{algpseudocode}
\usepackage{array,multirow,makecell}\setcellgapes{1pt}
\usepackage[table]{xcolor}
\usepackage{setspace}

\newcolumntype{R}[1]{>{\raggedleft\arraybackslash }b{#1}}
\newcolumntype{L}[1]{>{\raggedright\arraybackslash }b{#1}}
\newcolumntype{C}[1]{>{\centering\arraybackslash }b{#1}}
\makegapedcells

\renewcommand\IEEEkeywordsname{Index Terms}

\begin{document}
\pagestyle{empty}

\title{\huge Sum-rate Maximizing in Downlink Massive MIMO Systems with Circuit Power Consumption}
\author{\IEEEauthorblockN{Rami Hamdi $^{1,2}$ and Wessam Ajib $^{2}$} \\
\IEEEauthorblockA{$^{1}$ Department of Electrical Engineering, École de Technologie Supérieure  \\ rami.hamdi.1@ens.etsmtl.ca \\
$^{2}$ Department of Computer Science, Université du Québec à Montréal \\
ajib.wessam@uqam.ca }
}
\maketitle
\thispagestyle{empty}

\begin{abstract}

The downlink of a single cell base station (BS) equipped with large-scale multiple-input multiple-output (MIMO) system  is investigated in this paper. As the number of antennas at the base station becomes large, the power consumed at the RF chains cannot be anymore neglected. So, a circuit power consumption model is introduced in this work.  It involves that the maximal sum-rate is not obtained when activating all the available RF chains. Hence, the aim of this work is to find the optimal number of activated RF chains that maximizes the sum-rate. Computing the optimal number of activated RF chains must be accompanied by an adequate antenna selection strategy. First, we derive analytically the optimal number of RF chains to be activated so that the average sum-rate is maximized under received equal power. Then, we propose an efficient greedy algorithm to select the sub-optimal set of RF chains to be activated with regards to the system sum-rate. It allows finding the balance between the power consumed at the RF chains and the transmitted power. The performance of the proposed algorithm is compared with the optimal performance given by brute force search (BFS) antenna selection. Simulations allow to compare the performance given by greedy, optimal and random antenna selection algorithms.

\end{abstract}


\section{Introduction}

 A major challenge of designing future wireless networks is to support the huge increase in data traffic demand. Massive MIMO system is one important technology that will be adopted in the fifth generation (5G) of cellular networks~\cite{five1,five2,five3,five4}. It is based on the installation of large number of antennas at the Base Station (BS). The growth of the number of antennas involves quasi-orthogonality between the users' channels in consequence of the law of large numbers. Hence, a huge gain in spectral efficiency and energy efficiency can be obtained while using low complexity transmit and receive techniques such as Maximal Ratio Combiner (MRC), Maximal Ratio Transmission (MRT) and Zero Forcing (ZF)~\cite{over2,mrt,res1}.

Few works have previously investigated the optimization of resource allocation in massive MIMO systems. In~\cite{allo1}, the authors proposed an algorithm to compute the optimal power allocation among users in downlink massive MIMO systems with MRT precoder. However, a low complexity linear precoder is designed for downlink massive MIMO systems in~\cite{allo2}. The beamforming vector is computed jointly with optimal power allocation in order to maximize the sum-rate. The proposed precoder archives high performance near Regularized Zero Forcing (RZF) and better than MRT with no matrix inversion. In~\cite{allo3}, the authors investigate the uplink of multi-cell massive MIMO system with MRC receiver. An efficient algorithm is proposed to compute jointly the optimal training duration and power allocation. In~\cite{allo4}, the number of RF chains is assumed to be limited. Hence, a joint antenna selection and user scheduling strategy is introduced for downlink massive MIMO systems. As the large number of antennas involves the growth of the Channel State Information (CSI), the authors in~\cite{allo5} assume limited CSI at the BS and efficient antenna selection algorithm is proposed. In~\cite{allo6}, the authors propose a polynomial time algorithm to select antennas with maximum SNR.

In massive MIMO systems, as the number of antennas is getting larger, the power consumed by the RF chains becomes not negligible~\cite{cir1,cir2,cir3,cir4}. Consequently, the maximal sum-rate is not obtained when activating all the RF chains. Hence, the optimal number of RF chains to be activated that maximizes the sum-rate must be derived. In~\cite{cir3}, the authors investigate the optimization of the power allocation and the number of transmit antennas based on the asymptotic approximation of the capacity. They extend their work in~\cite{cir4} to multi-users and imperfect CSI case assuming MRT precoding. But, the optimization of the power allocated among users is still based on the average capacity over channel realizations. It will be more interesting if the optimization is based on the instantaneous throughput. As results, the optimal number of transmit antennas should be derived jointly with antenna selection strategy depending of the channel gain coefficients. In this work, we derive first analytically the optimal number of activated RF chains that maximizes the sum-rate averaged over channel realizations under the assumption of equal received power. Then, an efficient greedy algorithm is proposed to compute the close-to-optimal number of RF chains and to select the best subset of antennas that maximizes the sum-rate. The proposed algorithm allows to find the balance between the power consumed at the RF chains and the output transmitted power. Monte-Carlo simulations allow to compare the performance of the proposed algorithm with random and brute force search (BFS) optimal antenna selection.

In this paper, all power variables are assumed to be unitless since they are normalized by the average noise power. $\lfloor.\rfloor$ denotes the floor function, $\lceil.\rceil$ denotes the ceiling function, $\text{diag}(\mathbf{p})$ is a diagonal matrix whose diagonal entries are the elements of the vector $\mathbf{p}$, $(x)^+$ denotes $\text{max}(0,x)$, $(.)^H$ represents the Hermitian matrix,$\mathbf{C}_{N}^{S}$ is a binomial coefficient defined as $\mathbf{C}_{N}^{S}=\frac{S!}{(N-S)! N!}$ and $\mathbf{Tr}\{.\}$ denotes the trace of a square matrix.

The rest of the paper is organized as follows. In Section II, the system model is presented. In Section III, the problem is formulated. In Section IV, the optimal number of RF chains that maximizes the average sum-rate is analytically calculated under the assumption of equal received power between users.  Then, an efficient greedy algorithm is proposed to compute the number of RF chains that maximizes the sum-rate and to select the best antennas in Section V. The computational complexity of the proposed algorithm is calculated and compared to the optimal BFS algorithm in Section VI. Numerical and simulation results are shown and discussed in Section VII. Finally, we conclude and discuss the main results in Section VIII.

\section{System Model}

We consider the downlink of a single cell massive MIMO systems. The base station (BS) is equipped with a large number of antennas $N$ serving $K$ single-antenna users with $N \gg K$. The channel gain is represented by complex matrix $\bm{G}=\bm{D}^{1/2} \bm{H}$. The small scale fading is defined by complex matrix $\bm{H}=[\mathbf{h_1}, \mathbf{h_2}, ... ,\mathbf{h_K}]$, where $\mathbf{h_k} \in \mathbb{C}^{1\times N}$ is the $k^{th}$ channel vector for user $k$, is assumed to be quasi-static Gaussian independent and identically distributed (i. i. d.) slow fading channel. Considering only path loss, the large scale fading is defined by the vector $\mathbf{d}=[d_1^{-\alpha} d_2^{-\alpha} ... d_K^{-\alpha}]$, where $\alpha$ is the path loss exponent and $d_k$ the distance between the BS and user $k$, $\bm{D}$ is a diagonal matrix defined as $\bm{D}=\text{diag}(\mathbf{d})$. The noise is assumed to be additive Gaussian white (AWGN) random variable with zero mean and unit variance. The vector $\mathbf{p}=[p_1 p_2 ... p_K]$ is assumed to be the distribution of power among users. We consider that BS knows perfectly the CSI. We consider the Zero Forcing (ZF) as beamforming strategy because it cancels the inter-user interference and it achieves high performance~\cite{over2,mrt,res1}. The beam forming matrix is expressed as

\begin{equation}
 \label{eq:1}
   \bm{W}= \frac{\bm{H}^{H} (\bm{H} \bm{H}^{H})^{-1}}{\eta},
\end{equation}
where $\eta$ the normalization factor defined as $\eta=\sqrt{\mathbf{Tr}\{(\bm{H} \bm{H}^{H})^{-1}\}}$.

Hence, the received downlink vector of signals is expressed as

\begin{equation}
 \label{eq:2}
   \mathbf{y}= \frac{\bm{D}^{\frac{1}{2}}}{\sigma \eta} \text{diag}(\sqrt{\mathbf{p}}) \mathbf{s}+\mathbf{n},
\end{equation}
where $\mathbf{s}$ is the vector of transmitted data symbol and $\mathbf{n}$ is the vector of AWGN. $\sigma=\sqrt{\frac{\mathbf{Tr}\{\bm{D}\}}{K}}$ is a normalization factor.

The signal received by user $k$ can be written as

\begin{equation}
 \label{eq:3}
   y_k= \sqrt{p_k} \frac{d_k^{-\frac{\alpha}{2}}}{\sigma \eta} s_k +n_k.
\end{equation}

The received $SINR$ for user $k$ is expressed as

\begin{equation}
 \label{eq:4}
   {SINR}_k=\frac{d_k^{-\alpha} p_k}{\sigma^{2} \eta^{2}}.
\end{equation}

Hence, the sum-rate is expressed as

\begin{equation}
 \label{eq:60}
   R=\sum_{k=1}^{K} \log_2\left(1+\frac{d_k^{-\alpha} p_k}{\sigma^{2} \eta^{2}}\right).
\end{equation}

Since the number of antennas is large in the considered system, we assume a non-negligible circuit power consumption~\cite{cir1,cir2}. Let $p_c$ denotes the fixed power consumed at each activated RF chain (Digital to Analog Converter (DAC), mixer, frequency synthesizer, filter) and $p_{max}$ the maximal available power at the BS. Hence, the circuit power consumption constraint can be expressed as

\begin{equation}
\label{eq:7}
p_{out}+\sum_{n=1}^{N}\alpha_n . p_c \leq p_{max},
\end{equation}
where $\alpha_n$ is an antenna coefficient that is set to 1 if antenna $n$ is activated and to 0 otherwise and $p_{out}$ is the output transmitted power given by

\begin{equation}
\label{eq:8}
p_{out}=\sum_{k=1}^{K}p_k.
\end{equation}

It is to be noted that $\lfloor p_{max}/p_c\rfloor$ represents the maximum number of RF chains that can be supported (assuming no transmission power) by the system due to the circuit power constraint.

\section{Problem Formulation}

The circuit power consumption involves that the maximal achieved sum-rate is not obtained when activating all RF chains. Hence,  the optimal number of RF chains $\sum_{n=1}^{N}\alpha_n$ that maximizes the sum-rate should be derived. Moreover, the aim is to find the optimal balance between the power consumed by the RF chains $\sum_{n=1}^{N}\alpha_n . p_c$ and the transmitted power $p_{out}$. The number of RF chains should be optimized jointly with adequate antenna selection strategy. The transmitted power must be allocated optimally among users. Hence, the main problem can be formulated as

\begin{equation}
\label{eq:9}
\begin{aligned}
& \underset{\mathbf{p},\alpha_n:n=1..N}{\text{maximize}}
& & R= \sum_{k=1}^{K} \log_2\left(1+\frac{d_k^{-\alpha} p_k}{\sigma^{2} \eta^{2}(\alpha_1, \alpha_2, ... , \alpha_n)}\right) \\
& \text{subject to}
& & \sum_{k=1}^{K}p_k+\sum_{n=1}^{N}\alpha_n . p_c \leq p_{max}, \\
&&& \sum_{n=1}^{N}\alpha_n \geq K, \\
&&& \alpha_n \in \{0,1\},n=1..N.
\end{aligned}
\end{equation}

It is to be highlighted that the term $\eta^{2}$ in the objective function depends on the fading coefficients of the selected antennas. So, an antenna selection strategy is necessary to find the optimal number of RF chains that maximizes the sum-rate. Consequently, the problem becomes combinatorial with exponential complexity growth in $N$. It is known that, after selecting the best antenna, water filling is the optimal strategy for power allocation among users~\cite{wf}.

\section{Equal Received Power}

In this section, equal received power between users is assumed and the sum-rate averaged over channel realizations is approximated. Then, we derive analytically the optimal number of RF chains that maximizes the sum-rate averaged over channel realizations. The number of RF chains is given by

\begin{equation}
 \label{eq:10}
   S=\sum_{n=1}^{N}\alpha_n.
\end{equation}

Note that since the sum-rate is averaged over the channel realizations, the antenna selection is meaningless. Since equal received power between all users is assumed, the transmitted power allocated for user $k$ can be expressed as

\begin{equation}
 \label{eq:61}
   p_k=\frac{p_{max}-S.p_c}{K} \frac{\sigma^{2}}{d_k^{-\alpha}}.
\end{equation}

The term $\eta^{2}$ is approximated in~\cite{over2,mrt,res1,chann}, when $K,S \longrightarrow \infty$

\begin{equation}
 \label{eq:11}
 \frac{1}{\eta^{2}}=\frac{1}{\mathbf{Tr}\{(\bm{H} \bm{H}^{H})^{-1}\}} \longrightarrow \frac{S}{K}-1.
\end{equation}

This approximation is validated by simulations in Section VII. Hence, the sum-rate averaged over the channel realizations can be expressed as

\begin{equation}
 \label{eq:11}
   \overline{R}=K.\log_2\left(1+\frac{(p_{max}-S.p_c)(S-K)}{K^2} \right).
\end{equation}

In consequence, the optimal number of RF chains that maximizes the average sum-rate over channel realizations can be derived

\begin{equation}
 \label{eq:12}
   S^{*}=
   \begin{cases}
\lfloor \phi\rfloor & \text{if~} \overline{R}(\lfloor \phi\rfloor) > \overline{R}(\lceil \phi\rceil)  \text{~or~} \lfloor \phi \rfloor=\lfloor p_{max}/p_c\rfloor \\
\lceil \phi\rceil & \text{otherwise}
\end{cases}
\end{equation}
where
\begin{equation}
 \label{eq:13}
  \phi=\frac{p_{max}+K.p_c}{2 p_c} < p_{max}/p_c.
\end{equation}

Finally, the power allocated for each user $k$ is given by

\begin{equation}
 \label{eq:14}
   p_k=\frac{p_{max}-S^{*}.p_c}{K} \frac{\sigma^{2}}{d_k^{-\alpha}}.
\end{equation}

However, the number of RF chains that optimizes the instantaneous sum-rate (for one channels realization) should be derived. It depends of the coefficients of the selected antennas. In consequence, the number of RF chains should be optimized jointly with adequate antenna selection strategy which will be the focus of the next section.

\section{Efficient Greedy Algorithm}

The aim of this work is to find the optimal number of RF chains and have an antenna selection that maximizes the sum-rate. The number of RF chains must be derived jointly with adequate antenna selection strategy. The antenna selection is a combinatorial problem with exponential growth in $N$. Moreover, the optimal antenna selection can be done with high complexity brute-force search (BFS) algorithm. So, we propose a low complexity greedy iterative algorithm that determines the number of RF chains maximizing the instantaneous sum-rate. At each iteration for a fixed number of RF chains, the best antenna $n^{*}$ that maximizes the sum-rate is determined among the set of non selected antennas $\Lambda$. Since, the cost function $\eta^{2}$ is infinite for $S<K$ as the ZF Precoder does not exist, we choose the first $K$ antennas by optimal BFS. The best antenna should minimize the term $\eta^{2}$

\begin{equation}
 \label{eq:15}
   \text{argmax}_{\Lambda} R =\text{argmin}_{\Lambda} \eta^{2}(\alpha_1, \alpha_2, ... , \alpha_N)
\end{equation}

Hence, the best antenna at each iteration can be derived as

\begin{equation}
 \label{eq:63}
   n^{*}=\text{argmin}_{\Lambda} \eta^{2}(\alpha_1, \alpha_2, ... , \alpha_N)
\end{equation}

As the selected antennas are given, the optimal strategy for power allocation among users is given by Water filling~\cite{wf}

\begin{equation}
 \label{eq:16}
   p_k=\left( \frac{1}{\ln(2) \mu} -\frac{\sigma^{2} \eta^{2}}{d_k^{-\alpha}}  \right)^{+},
\end{equation}
where $\mu$ is the Water level.

The convergence of the proposed algorithm is obtained when the instantaneous sum-rate starts decreasing ($R_S < R_{S-1}$). The convergence point exists since the sum-rate is a concave function, the second order derivative is negative $\frac{d^2 R}{dS^2} < 0$. Hence, the proposed algorithm allows to determine the number of RF chains, the selected antennas, the power allocated among users and the achieved sum-rate.

\begin{algorithm}[h]
\caption{Greedy Algorithm}
\begin{algorithmic}[1]
\State Initialization
\State $\alpha_n \gets 0, n=1:N$
\State $\Lambda \gets \{n, n=1:N\}$
 \For{$S=K+1:\lfloor p_{max}/p_c\rfloor$}
\State Select the  best antenna $n^{*}$ from $\Lambda$ that minimizes $\eta^{2}$
\State $\alpha_{n^{*}} \gets 1$
\State $\Lambda \gets \Lambda\setminus \{n^{*}\}$
\State  $\sum_{k=1}^{K} \left( \frac{1}{\ln(2) \mu} -\frac{\sigma^{2} \eta^{2}}{d_k^{-\alpha}}  \right)^{+}=p_{max}-S.p_c$ \Comment{find $\mu$ by bisection method}
\State $p_k \gets \left( \frac{1}{\ln(2) \mu} -\frac{\sigma^{2} \eta^{2}}{d_k^{-\alpha}} \right)^{+}, k=1:K$, power allocation
\State Break If $R_S < R_{S-1}$
\EndFor
\end{algorithmic}
\end{algorithm}

\section{Complexity Analysis}

In this section, the computational complexity order for both optimal brute force antenna selection and the proposed greedy algorithm are calculated. For the brute-force search optimal algorithm, the number of possible combinations of sets of the antennas is

\begin{equation}
 \label{eq:17}
  \sum_{S=1}^{\lfloor p_{max}/p_c\rfloor} \mathbf{C}_{N-S+1}^{S}
\end{equation}

For the proposed greedy algorithm, the number of combinations of sets of the antennas is given by

\begin{equation}
 \label{eq:71}
  \sum_{S=1}^{\lfloor p_{max}/p_c\rfloor} N-S+1
\end{equation}

For each selected set of antennas, the coefficient $\eta^{2}$ is calculated. So, we need one matrix multiplication and one matrix inversion. The complexity order of the matrix multiplication is $O(S K^2)$ and the complexity order for the matrix inversion is $O(K^3)$.

Hence, the computational complexity order of the BFS algorithm is given by

\begin{equation}
 \label{eq:18}
  C_{BFS}=O\left(\sum_{S=1}^{\lfloor p_{max}/p_c\rfloor} \mathbf{C}_{N-S+1}^{S} (S K^2+K^3)\right)
\end{equation}
and the computational complexity order of the greedy algorithm is given by

\begin{equation}
 \label{eq:19}
 \begin{aligned}
  C_{greedy} & = O\left(\sum_{S=1}^{\lfloor p_{max}/p_c\rfloor} (N-S+1) (S K^2+K^3)\right)\\
 &  = O( \lfloor p_{max}/p_c\rfloor K^2 (K (N+1) \\
 &  + \frac{1}{2} (N+1-K) (\lfloor p_{max}/p_c\rfloor +1)\\
 &  + \frac{1}{6} (\lfloor p_{max}/p_c\rfloor +1) (2 \lfloor p_{max}/p_c\rfloor +1)))
 \end{aligned}
\end{equation}

Hence, the optimal antenna selection can be obtained with very high complexity. However, the proposed efficient algorithm is polynomial time. The complexity order for both algorithms are presented in Table~\ref{tab1}.

\begin{table}[h]
\caption{Computational Complexity of the proposed greedy and BFS algorithms.}
\label{tab1}
\begin{tabular}{|L{4cm}|L{1.5cm}|L{1.5cm}|}
\hline \rowcolor{lightgray}  & BFS &  Greedy  \\
\hline  Complexity order & $C_{BFS}$ & $C_{greedy}$ \\
\hline  \tiny{$N=64, K=10, \lfloor p_{max}/p_c\rfloor=32$}  & $O(8.10^{16})$ & $O(6.10^{5})$ \\
\hline  \tiny{$N=128, K=10, \lfloor p_{max}/p_c\rfloor=128$}  & $O(3.10^{30})$ & $O(4.10^{7})$ \\
\hline
\end{tabular}
\end{table}

\section{Numerical Results}

In this section, we present numerical results to show the performance of the proposed greedy algorithm compared to random antenna selection and optimal BFS algorithm. Monte-Carlo simulations are done to show the optimized number of RF chains and the maximal achieved sum-rate.

Simulation parameters are summarized  in Table~\ref{tab2}. The BS is equipped with 256 antennas serving 10 users.

\begin{table}[h]
\caption{System Parameters.}
\label{tab2}
\begin{tabular}{|R{1cm}|C{4cm}|L{2cm}|}
\hline \rowcolor{lightgray} Symbol & Description &  Value  \\
\hline  $K$ & number of users & 10  \\
\hline  $N$ & number of antennas & 256  \\
\hline  $P_c$ & fixed power RF chains & 0.05 \\
\hline  $\alpha$ & path loss exponent & 3.7 \\
\hline   & cell radius & 500 m \\
\hline
\end{tabular}
\end{table}

\begin{figure}[h]
 \centerline{\includegraphics [width=9cm]  {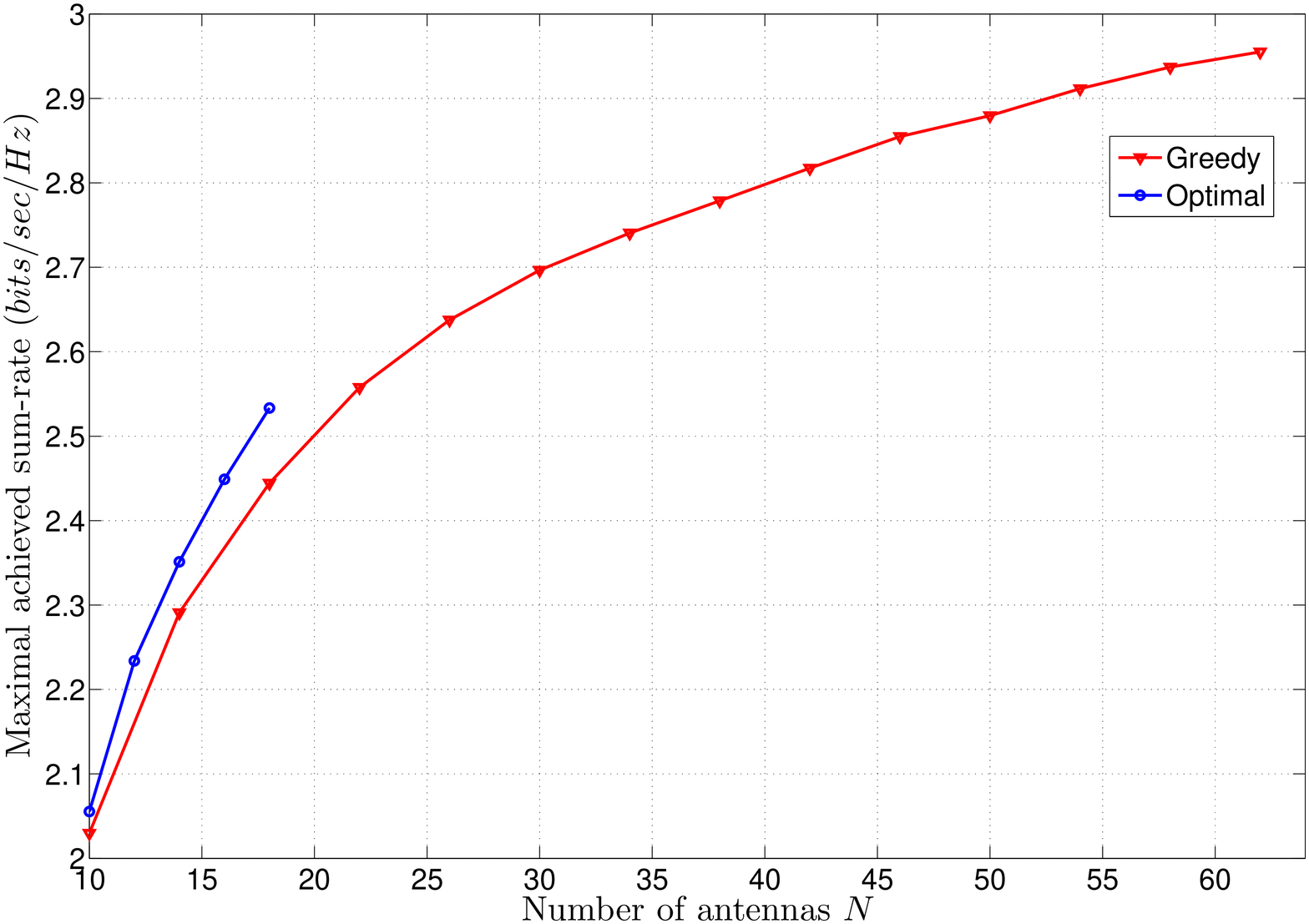}}
\caption{Maximal achieved sum-rate under optimal and greedy antenna selection ($K=3, p_{max}=1$).}
\label{fig1}
\end{figure}

Fig.~\ref{fig1} shows the maximal achieved sum-rate in function of the number of antennas $N$ under the proposed greedy algorithm and optimal antenna selection. Due to the high complexity of BFS algorithm, simulation results for optimal antenna selection are presented for limited number of antennas $N$, three users $K=3$ and $p_{max}=1$. It is clear that the achieved sum-rate increases as the number of antennas increases since it offers more diversity to the transmitter. The difference between the performance of the low complexity proposed greedy algorithm and the optimal antenna selection strategy is small.

\begin{figure}[h]
 \centerline{\includegraphics [width=9cm]  {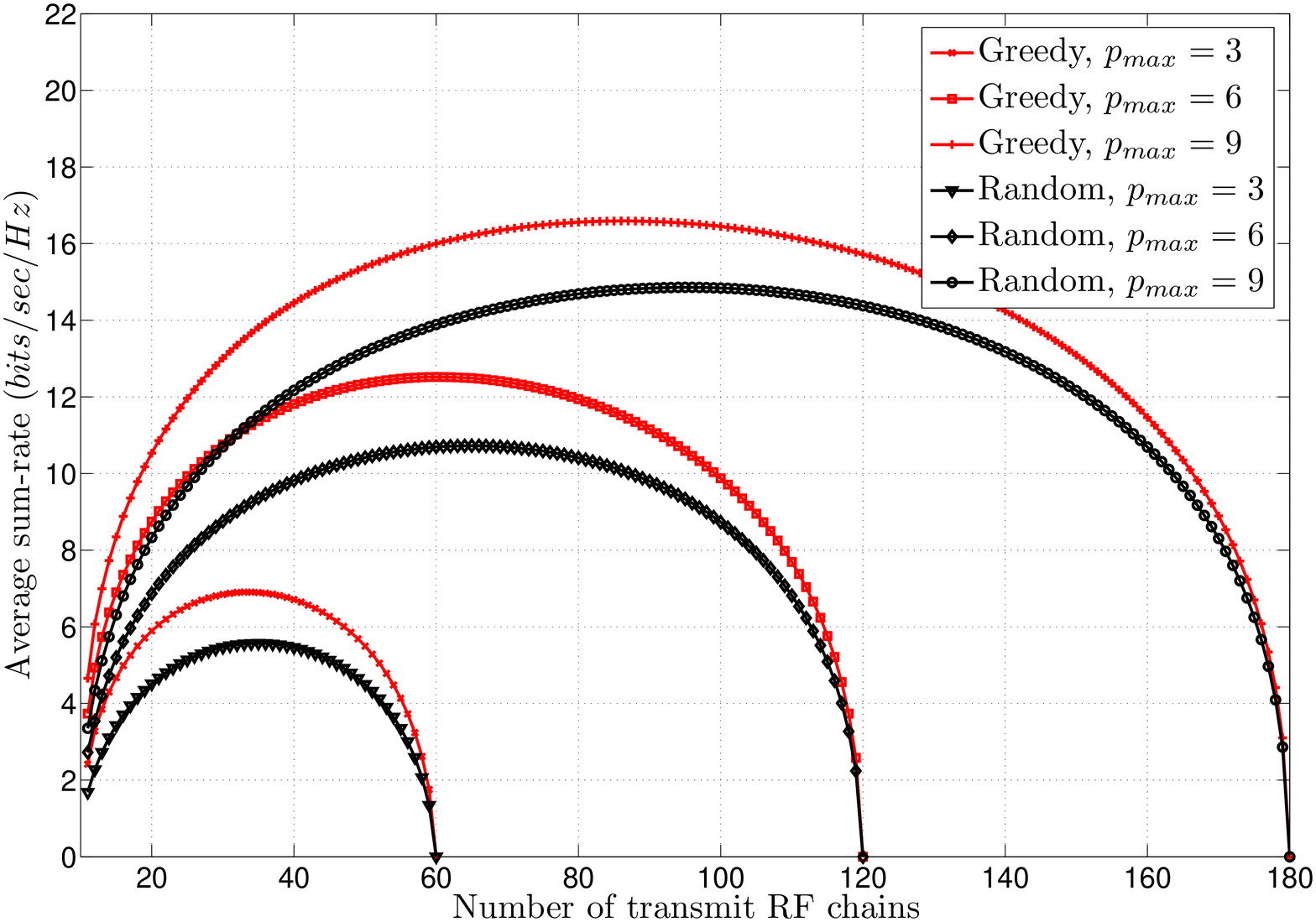}}
\caption{Average sum-rate under random and greedy antenna selection.}
\label{fig2}
\end{figure}

In Fig.~\ref{fig2}, we see the achieved average sum-rate in function of the number of RF chains under the proposed greedy algorithm and random antenna selection for different values of total available power at the BS. First, the obtained results confirm the hypothesis that the maximal achieved sum-rate is not obtained when activating all RF chains. Then, the performance of the proposed greedy algorithm is significantly higher than the random antenna selection because the greedy algorithm allows to select the best antenna at each iteration.

\begin{figure}[h]
 \centerline{\includegraphics [width=9cm]  {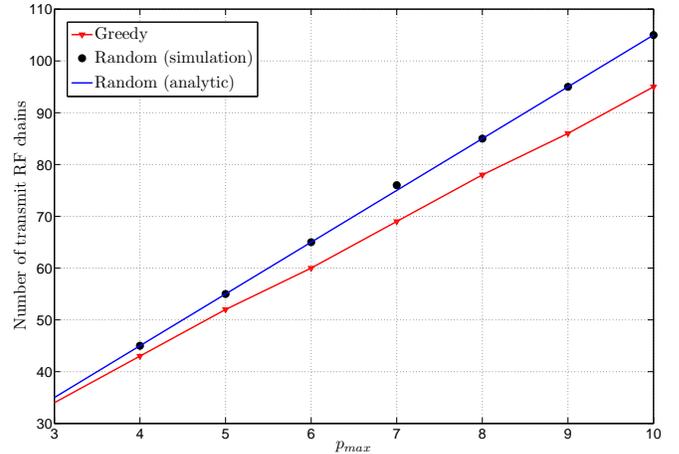}}
\caption{Number of activated transmit RF chains under random and greedy antenna selection.}
\label{fig3}
\end{figure}

Fig.~\ref{fig3} shows the optimal number of RF chains to be activated that maximizes the average sum-rate in function of the maximal available power at the BS. We observe that the analytic expression of the optimal number of RF chains obtained under equal received power and random antenna selection fits with simulation results. Then, it is clear that the optimized number of RF chains obtained by the proposed greedy algorithm is less than the number of RF chains obtained under random antenna selection due to the adequate antenna selection strategy and the power is optimally distributed among users. So, the proposed algorithm allows to use less number of transmitted RF chains and more power is available at the base station for transmitting. Hence, it allows to increase the achieved sum-rate.

\begin{figure}[h]
 \centerline{\includegraphics [width=9cm]  {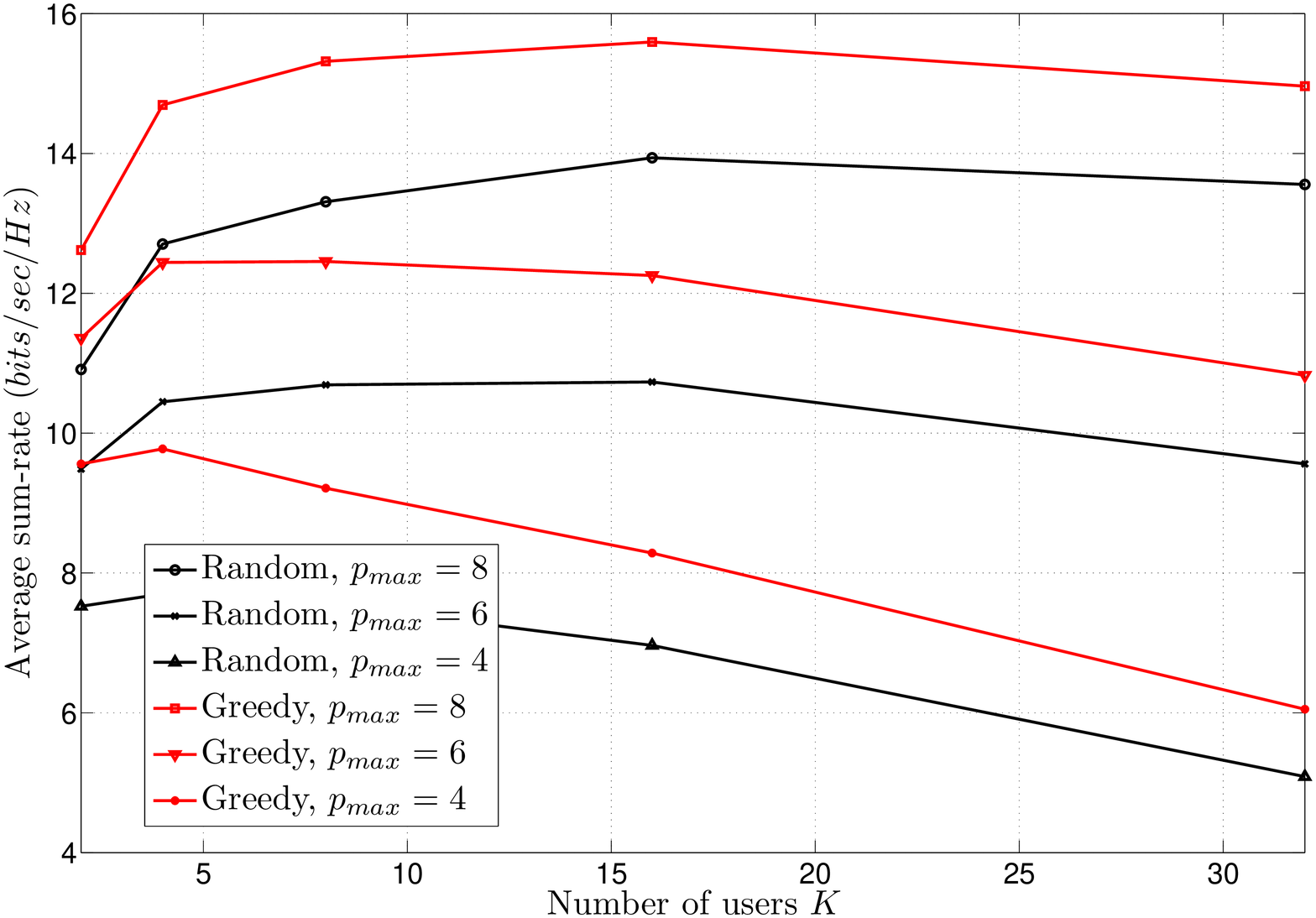}}
\caption{Impact of the number of user $K$ on the system performance.}
\label{fig4}
\end{figure}

Now, we investigate the impact of the number of users $K$ on the system performance. In Fig.~\ref{fig4}, we see the maximal achieved sum-rate in function of the number of users under the proposed greedy algorithm and random antenna selection. So, the optimal number of users that maximizes the sum-rate can be observed for different values of that maximal available power $p_{max}$ at the BS. The proposed greedy algorithm allows to transmit with less number of RF chains than random antenna selection. So, there is more available power for transmitting $p_{out}$. Hence, more users can be served under the proposed greedy algorithm and the maximal achieved sum-rate increases significantly.

\begin{figure}[h]
 \centerline{\includegraphics [width=9cm]  {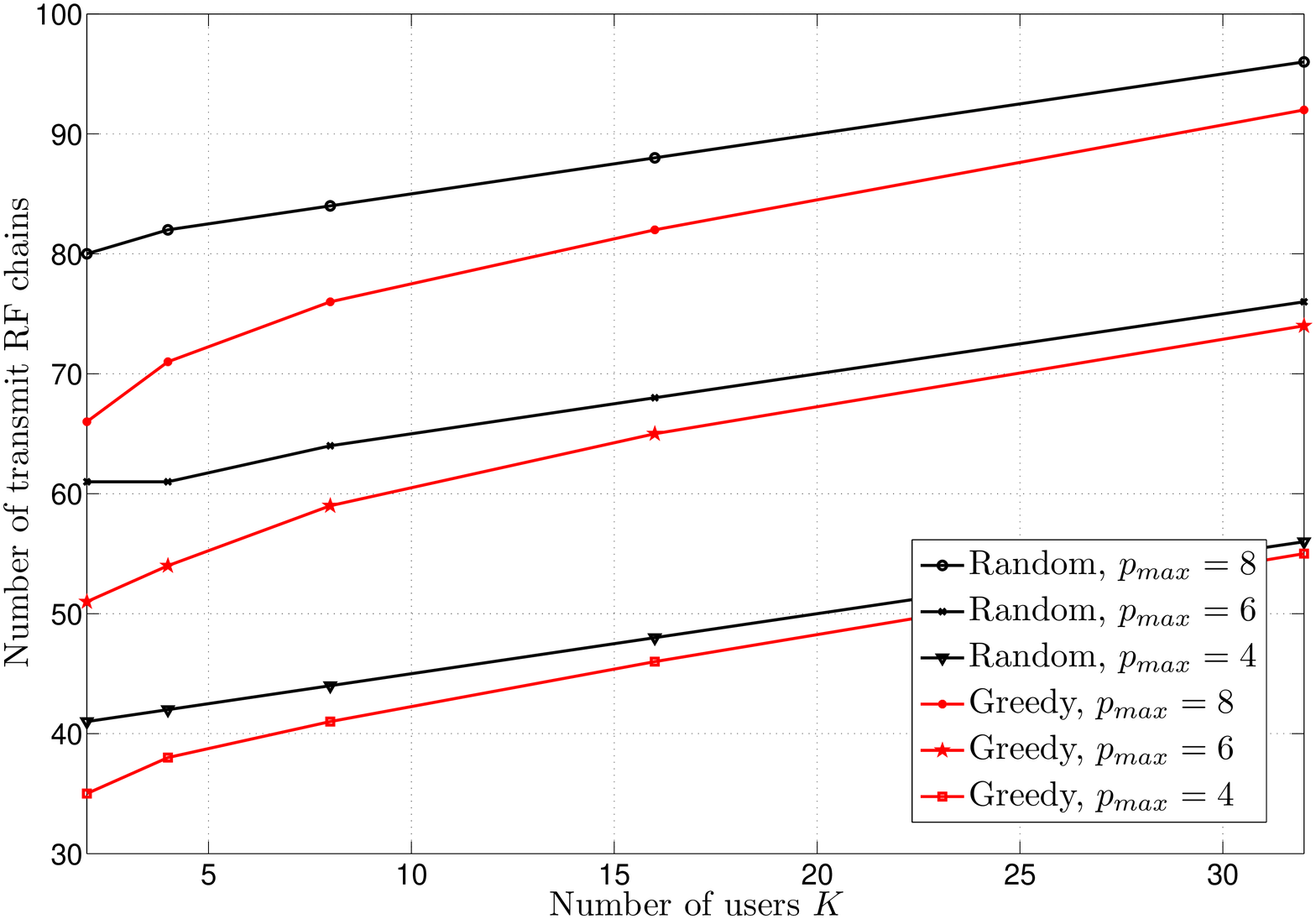}}
\caption{Number of activated transmit RF chains in function of the number of users $K$.}
\label{fig5}
\end{figure}

Finally, Fig.~\ref{fig5} shows the optimized number of RF chains in function of the number of users $K$ for different values of the maximal available power at the BS. The number of RF chains increases when increasing the number of users $K$ since we need more RF chains for transmitting. Also, as the maximal available power $p_{max}$ increases, there is more flexibility to use more transmit RF chains. It is clear that the computed number of RF chains under the greedy algorithm is less than random antenna selection. In consequence, there is more available power for transmitting and the system performance is ameliorated.

\section{Conclusion}

The downlink massive MIMO system with non negligible circuit power consumption is investigated in this paper. Since this work considers the circuit power consumption, the maximal sum-rate is not obtained when activating all the available RF chains. For this reason, we focus in this paper on optimizing the number of activated RF chains that maximizes the sum-rate. So, we derive analytically the optimal number of activated RF chains that maximizes the average sum-rate under equal received power. However, the number of RF chains that maximizes the sum-rate should be optimized jointly with adequate antenna selection. Hence, an efficient greedy iterative algorithm that optimizes the number of activated RF chains is proposed. It allows to find the balance between the power consumed at the RF chains and the transmitted power. Simulations are done to compare performance of the proposed algorithm to random and optimal antenna selection. Numerical results show the efficiency of the proposed algorithm.

\end{document}